\documentclass[10pt,aps,twocolumn,prd,superscriptaddress,floatfix]{revtex4-1}

\usepackage[utf8]{inputenc}

\usepackage{mathtools}
\usepackage{amsfonts}
\usepackage{mathrsfs}
\usepackage{bbm}
\usepackage{dsfont}

\usepackage{graphicx}
\usepackage{xcolor}

\usepackage{placeins}

\usepackage{xspace}
\usepackage{siunitx}
\usepackage{hyperref}

\usepackage{xifthen}

\setkeys{Gin}{width=0.48\textwidth}
\def\App#1{App.~\ref{#1}}

\graphicspath{{./figures/}}

\renewcommand{\vec}[1]{\mathbf{#1}}

\newcommand{\imag}{\mathrm{i}}
\newcommand{\contourC}{\mathcal{C}}

\DeclareMathOperator{\e}{e}
\DeclareMathOperator{\sgn}{sgn}
\DeclareMathOperator{\sgnc}{\sgn_\mathcal{C}}
\DeclareMathOperator{\deltac}{\delta_\mathcal{C}}

\DeclareMathOperator{\Tr}{Tr\!}
\DeclareMathOperator{\T}{{\cal T}}
\DeclareMathOperator{\Tc}{{\T}_{\!\!\contourC}}

\def\Fig#1{Figure~\ref{#1}}
\def\fig#1{Figure~\ref{#1}}
 
\def\Eq#1{Eq.~(\ref{#1})}
\def\eq#1{(\ref{#1})}

\def\eqref#1{(\ref{#1})}

\def\sec#1{Section~\ref{#1}}
\def\app#1{Appendix~\ref{#1}}
\def\App#1{Appendix~\ref{#1}}
\newcommand{\gettitle}{Flowing with the Temporal Renormalisation Group}

\hypersetup{
	pdftitle={\gettitle},
	pdfauthor={Corell, Cyrol, Heller, Pawlowski},
	pdfkeywords={time evolution} {scalar theory}
		{equilibrium} {nonthermal},
	bookmarksopen=true,
	bookmarksopenlevel=2,
	bookmarksnumbered=true
}

\begin{document}

\title{\gettitle}

\author{Lukas Corell}
\affiliation{Institut f\"ur Theoretische Physik,
	Universit\"at Heidelberg, Philosophenweg 16,
	69120 Heidelberg, Germany
}

\author{Anton K. Cyrol}
\affiliation{Institut f\"ur Theoretische Physik,
	Universit\"at Heidelberg, Philosophenweg 16,
	69120 Heidelberg, Germany
}

\author{Markus Heller}
\affiliation{Institut f\"ur Theoretische Physik,
	Universit\"at Heidelberg, Philosophenweg 16,
	69120 Heidelberg, Germany
} 

\author{Jan M. Pawlowski}
\affiliation{Institut f\"ur Theoretische Physik,
	Universit\"at Heidelberg, Philosophenweg 16,
	69120 Heidelberg, Germany
}
\affiliation{ExtreMe Matter Institute EMMI,
	GSI, Planckstr. 1,
	64291 Darmstadt, Germany
}

\begin{abstract}
  We discuss the far-from-equilibrium evolution of $\phi^3$-theory in
  $1+1$ dimensions with the temporal functional renormalisation group
  \cite{Gasenzer:2007za, Gasenzer:2010rq}. In particular, we show that
  this manifestly causal approach leads to novel one-loop exact
  equations for fully dressed correlation functions. Within this
  setup, we numerically compute the dynamical propagator. Its
  behaviour suggests self-similarity far from equilibrium in a
  restricted momentum regime. We discuss the scaling exponents for our
  solution, as well as the numerical satisfaction of energy and
  particle number conservation. We also derive a simple exact
  representation of the expectation value of the energy-momentum
  tensor solely in terms of the propagator.
\end{abstract}

\maketitle

\section{Introduction}
\label{sec:introduction}

The dynamics of quantum systems far from equilibrium has received much
attention in the past decades.  The theoretical resolution of these
processes is crucial for the understanding of systems reaching from
early universe cosmology and heavy ion collisions to the dynamics of
ultracold atom clouds in table top experiments. This has triggered
many theoretical developments ranging from lattice to diagrammatic
approaches, for reviews see e.g.~\cite{Nowak:2013juc, Berges:2015kfa,
  Schmied:2018mte,Berges:2020fwq}.

In the present work, we apply and develop further a functional
renormalisation group (fRG) approach,
\cite{Wetterich:1992yh,Ellwanger:1993mw,Morris:1993qb,Dupuis:2020fhh}, to
non-equilibrium physics put forward in \cite{Gasenzer:2007za,
  Gasenzer:2010rq} based on a temporal cutoff. For related
developments in the context of cosmology see also
\cite{Pietroni:2008jx}, for non-equilibrium fRG applications with a
standard momentum cutoff see e.g.\ \cite{2007PhRvL..99o0603J,
  Berges:2008sr, 2010PhRvB..81s5109J, Canet:2009vz,
  2012PhRvB..85h5113K, Canet:2011ez, Berges:2012ty,
  2013PhRvL.110s5301S, Mesterhazy:2013naa, 2014PhRvB..89m4310S,
  Canet:2014dta, Mathey:2014xxa, Mesterhazy:2015uja,
  Chiocchetta:2016waa, Duclut:2016jct, Tarpin:2017uzn, Tarpin:2018yvs,
  2018arXiv180210014S,Huelsmann:2020xcy,Nagy:2020bal,Wilkins:2021wde}.

The formulation of the cutoff in momentum space leads to a
modification of local conservation laws reaching from conserved
charges to gauge theories. While this modification is well-captured
within modified symmetry identities, the control of such modifications
is even more important for the dynamics of a system. There, a
violation of conservation laws can cause secularities and hence a
breakdown of the approach at hand. This has triggered the development
of an fRG approach for quantum dynamics based on a (temporal) cutoff
of the Schwinger-Keldysh contour \cite{Gasenzer:2007za,
  Gasenzer:2010rq}. This cutoff
simply suppresses the time evolution of a system beyond the cutoff
time $\tau$. Trivially, the temporal flow of the system with this
cutoff time is manifestly causal, and the flow equation captures the
time evolution of the system at $t=\tau$. Apart from its manifest
causality, its formulation in position space preserves all local
conservation laws including gauge symmetries. In summary, it is the
manifest preservation of causality and of local symmetries which is at
the root of the temporal fRG (t-fRG) approach.

In the present work, we further develop the t-fRG approach as well as
applying it to the dynamics of a $\phi^3$\nobreakdash-theory in 1+1
dimensions. Such a theory is an ideal test case for the present
approach, but it is also of interest for extending the
far-from-equilibrium universality known from $\phi^4$-interactions
(relativistic and non-relativistic, e.g.\cite{PineiroOrioli2015}) and
gauge theories to the $\phi^3$\nobreakdash-theory. Furthermore, it is
interesting with regard to non-Abelian gauge theories that have both,
microscopic three-field and four-field vertices.

In \sec{sec:t-fRG} we briefly review the temporal fRG and
develop it further. This leads us to integrated flow equations, which
are novel one-loop exact dynamical relations for the full
unregularised correlation functions. In \sec{sec:scalar_field} we
apply the approach to the $\phi^3$\nobreakdash-theory in 1+1~dimensions
and compute the dynamics of the propagator. We also derive simple
relations for the expectation value of the energy-momentum tensor solely
in terms of the propagator. We discuss the self-similar scaling
behaviour of the propagator as well as the numerical satisfaction of
energy and particle number conservation. Our results are briefly
summed up in \sec{sec:conclusion}.


\section{Temporal functional renormalisation group}
\label{sec:t-fRG}

In this section we briefly review the temporal functional RG (t-fRG)
approach, for more details see~\cite{Gasenzer:2007za,
  Gasenzer:2010rq}. We discuss further formal developments that
are also important for the present numerical application. In
particular, we derive one loop exact functional relations for the full
(cutoff independent) correlation functions of a given quantum theory
by simply integrating the temporal cutoff parameter.

\subsection{Closed time path}
\label{sec:closed_time_path}

In the context of non-equilibrium phenomena, it is instructive to
employ a real-time formalism. We choose to work in the Schwinger-Keldysh
formalism, which was developed in~\cite{Schwinger1961a,
  Mahanthappa1962, Bakshi1963, Bakshi1963a, Keldysh1964,
  Keldysh1965}. In this section, we introduce this approach and
discuss its basic properties with regard to the time evolution of
correlation functions.

\begin{figure}[t]
	\centering
	\includegraphics[width=0.4\textwidth]{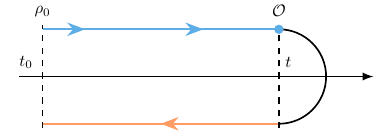}
	\caption{Closed time path used to compute the time dependent
          expectation value of the operator ${\cal O}$. ${\rho}_0$ is
          the density matrix at the initial time $t_0$. Note that the
          time path does not need to extend further than the latest
          operator insertion.}
	\label{fig:closed_time_path}
\end{figure}

At first, we consider the density matrix $\rho(t)$ which contains all
information on the state of a quantum system at a time $t$. Usually,
the density matrix is known at some initial time $t_{0}$ at which we
prepare our system. It could very well describe a system in thermal
equilibrium, where $\rho_{0}=\rho(t_{0})\propto\e^{-\beta H}$ with the
inverse temperature $\beta$ and the Hamiltonian $H$. Generally, it
describes any system and, in particular, interesting ones
far-from-equilibrium. The time evolution of the density matrix is
governed by the unitary time evolution operator $U$ and we write
\begin{align}
	{\rho}(t) = U(t,t_0) \, \rho(t_0) \,U(t_0,t) \,.
	\label{eq:rhot}
\end{align}
Typically, we are interested in the time evolution of an operator
expectation value $\langle{\cal O}\rangle(t)$, e.g. a correlation
function. For any operator, the expectation value at time~$t$ is given
as the trace over the density matrix and the operator
\begin{align}
	\langle {\cal O} \rangle (t) = \Tr \big[{\rho}(t) {\cal O}\big] \,.
\end{align}
To work out a more illustrative form of this expectation value, we use
\eqref{eq:rhot} and exploit that the trace is invariant under cyclic
permutations of the operators to find
\begin{align}
  \langle {\cal O} \rangle (t) = \Tr \big[{U}(t_0,t) \,{\cal O}\,
  {U}(t,t_0) {\rho}(t_0) \big] \,.
	\label{eq:ot}
\end{align} 
Reading the argument of the trace from right to left, the initial
density matrix is evolved in time from $t_0$ to $t$ where the operator
${\cal O}$ is inserted. Subsequently, there is a time evolution back to the
initial time. This operator ordering, which starts and ends at the
same time, directly suggests the term closed time path~(CTP) which is
used in the Schwinger-Keldysh formalism. A representation of the closed time
path for the expectation value $\mathcal{O}(t)$ is shown in
\fig{fig:closed_time_path}.

It is now possible to extend the closed time path to infinity by using
the unitarity of the time evolution operator. This is accomplished by
inserting $U(t,\infty)U(\infty,t)~=~\mathds{1}$ either to the left or
the right of the operator in \eqref{eq:ot}. We denote this extended
path with $\contourC$.  Depending on which side we extend the time
path, the operator is regarded as being placed on the forward or
backward branch of the CTP. Due to the construction, it is clear that
there is no difference for a single operator.

As soon as we insert multiple operators, it is crucial where each one
is placed. This is incorporated into the formalism by considering 
contour-time-ordered correlation functions. This time-ordering on the closed
time path is best understood as walking along the path in the
direction of the arrow in \fig{fig:closed_time_path}. Thus, 
time-ordering on the forward branch is the usual 
time-ordering, while on the backward branch, it is anti-time-ordering.
In particular, this means that every time on the backward part of the
contour is considered later than any time on the forward part.

Now, we define the generating functional for non-equilibrium
correlation functions
\begin{align}
  Z[J;{\rho}] = \Tr\bigg[{\rho}(t_0)\Tc\exp \bigg\lbrace \imag
  \int\displaylimits_{\mathrlap{\contourC(x)}}
  {\varphi}(x)J(x)	\bigg\rbrace \bigg] \,,
	\label{eq:partitionfunction}
\end{align}
where we introduced the notation
$\int_{\contourC(x)} = \int_{\contourC}\mathrm{d} x^{0}
\int_{\mathbb{R}^{d}} \mathrm{d}^d x$, the contour time ordering
operator $\Tc$ and the source $J$. Correlation functions are now
obtained as functional derivatives with respect to the source
\begin{align}
  \langle {\varphi}(x_1)\cdots{\varphi}(x_n) \rangle
  = (- \imag)^n\frac{ \delta^n Z[J;{\rho}]}{\delta J(x_1)\cdots\delta J(x_n)}
  \bigg\rvert_{J=0} \,, 
\end{align}
for more details see \cite{Gasenzer:2007za, Gasenzer:2010rq}.

\subsection{Flow equation}
\label{sec:flow_equation}

\begin{figure}[t]
	\centering
	\includegraphics[width=0.4\textwidth]{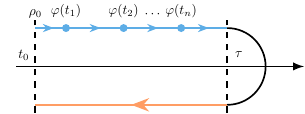}
	\caption{$n$-point function on the closed time path with the
          latest time $t_n<\tau$. Hence, all fluctuations up to $t_n$
          are included and we obtain the fully dressed correlation
          function.}
	\label{fig:closed_time_path_causality}
\end{figure}

Consider now a CTP that only extends to some finite (cutoff) time
$\tau$. The corresponding generating functional $Z_\tau[J;{\rho}]$
only sums over fluctuations up to $\tau$. Therefore, it contains no
sources for quantum fluctuations for times later than
$\tau$. Causality entails that all $n$-point functions derived from
$Z_\tau$ are the full correlation function as derived from
\eqref{eq:partitionfunction}, if all time arguments $x_i^0$ with
$i=1,...,n$ obey $x_i^0 \leq \tau$.  See
\fig{fig:closed_time_path_causality} for a graphical
representation.

\begin{figure}[t]
	\centering
	\includegraphics[width=0.4\textwidth]{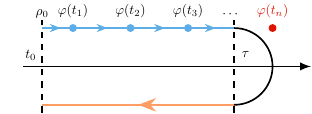}
	\caption{$n$-point function on the
          closed time path with the latest time $t_n>\tau$. Such
          fluctuations are not included in the generating functional
          $Z_\tau$. Thus, this correlator vanishes identically.}
	\label{fig:closed_time_path_causality_2}
\end{figure}

In turn, if at least one time argument of the n-point function is
larger than $\tau$, the correlator vanishes identically.
\Fig{fig:closed_time_path_causality_2} shows such a scenario.

We now turn to the effective action $\Gamma_\tau$, the generating
functional of 1PI correlation functions with time arguments smaller
than or equal to $\tau$. To that end, we consider the Schwinger
functional
\begin{align}
	W_\tau[J;{\rho}] = -\imag \ln{Z_\tau[J;{\rho}]} \,,
\end{align}
which inherits the causal properties from $Z_\tau$. Finally, we are
interested in moving the cutoff time $\tau$ to infinity. For deriving
a flow equation for the generating functionals, it is more convenient
to implemented the finite cutoff time via regulator terms as also used
for standard momentum cutoffs in the fRG. Hence, we add a cutoff term
to the classical action
\begin{align}\label{eq:Cutoff}
  \Delta S_\tau[\varphi]    = \frac{1}{2}
  \int\displaylimits_{\mathrlap{\contourC(x,y)}}
  \varphi(x)R_\tau (x,y) \varphi(y) \,.
\end{align}
We demand that the cutoff term \eq{eq:Cutoff} vanishes for both times
$x^0, y^0\leq \tau$ and suppresses all fluctuations for later times. The
respective regulator $R_\tau$ is specified later in
\sec{sec:causality}. Such a term implements the discussed causal properties.
The effective action is then obtained as the Legendre transformation
\begin{align}\nonumber 
  \Gamma_\tau[\phi;{\rho}] =& 
  W_\tau[J;{\rho}] -
    \int\displaylimits_{\mathrlap{\contourC(x)}} \phi(x)J(x) \\[1ex]
  &-\frac{1}{2} \int\displaylimits_{\mathrlap{\contourC(x,y)}}
    \phi(x)R_\tau (x,y) \phi(y) \,,
\label{eq:Gamma}\end{align}
with the field expectation value
$\phi(x)=\langle\varphi(x)\rangle$. Completely analogously to the
equilibrium case, we can derive a flow equation for the effective
action
\begin{align}
  \partial_\tau\Gamma_\tau[\phi]=\frac{1}{2}
  \int\displaylimits_{\mathrlap{\contourC(x,y)}}
  G_{\tau}(x,y)\,\partial_\tau R_{\tau}(x,y) \,.
    \label{eq:flow_equation}
\end{align}
A detailed derivation can be found in \cite{Gasenzer:2010rq}.  This
flow equation successively integrates out all relevant fluctuations
time-slice by time-slice, thus providing us with the time evolution of
the considered system. The regularised propagator $G_\tau$ is given by
\begin{align}
  \nonumber
  G_\tau(x,y) = -\imag \frac{\delta^2 W_\tau[J;{\rho}]}{
  \delta J(x) \delta J(y)} \bigg\rvert_{J=0} \,,
\end{align}
and is related to the 1PI two-point function by
\begin{align}\label{eq:DefofG}
  \imag G^{-1}_\tau(x,y) &= \big[\Gamma^{(2)}_\tau
                           +R_\tau\big](x,y) \,.
\end{align}
Flow equations for the 1PI correlation functions
$\Gamma^{{(n)}}$ are obtained by taking functional derivatives
with respect to the field expectation value
\begin{align}
  \Gamma_\tau^{(n)}(x_{1},\dots,x_{n}) &=
                                                \frac{\delta^{n}\Gamma_\tau}{
                                                \delta\phi(x_{1})\cdots\delta\phi(x_{n})}\,.
	\label{eq:1pi_correlation_functions}
\end{align}
Analogously to the flow equation in momentum space, this yields an
infinite hierarchy of coupled differential equations for the 1PI
correlation functions. To solve it, this hierarchy has to be
truncated.

\begin{figure}[t]
	\centering
	\includegraphics{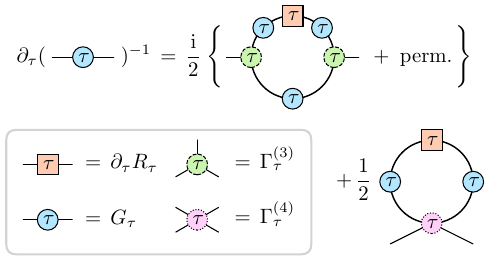}
	\caption{Diagrammatic representation of the flow equation of
          the two-point function. The orange square denotes
          $\partial_\tau R_\tau(x,y)$. Lines with blue circles
          represent the fully dressed propagator
          $G_\tau(x,y)~=~\imag\big[\Gamma^{(2)}_\tau+R_\tau\big]^{-1}(x,y)$. The
          green circle with three lines attached is
          $\Gamma^{(3)}_\tau$ and the pink circle with four legs is
          $\Gamma^{(4)}_\tau$, both also fully dressed. perm. stands
          for the permutation of the external legs.}
	\label{fig:flow_equation_diagrams}
\end{figure}

For now, we focus on the flow equation of the two-point function,
which is depicted in \fig{fig:flow_equation_diagrams}.  The full,
i.e. not truncated, flow reads
\begin{align}
  \partial_\tau \Gamma^{(2)}_\tau
  &(x,y) = 
    \frac\imag2 \int\displaylimits_{\mathrlap{\contourC(z_1,z_2)}}
    \Gamma^{(4)}_\tau (x,y,z_1,z_2) 
    \big( G_\tau \partial_\tau R_\tau G_\tau \big)(z_2,z_1) \nonumber\\[1ex]	
  -&\frac{1}{2} \int\displaylimits_{\mathrlap{\contourC(z_1,\dots,z_4)}}
     \Big\lbrace \Gamma^{(3)}_\tau(x,z_1,z_2) G_\tau(z_2,z_4)
     \Gamma^{(3)}_\tau(z_3,z_4,y) \nonumber\\[1ex]  
  &\qquad\quad\times \big.\big( G_\tau \partial_\tau R_\tau
    G_\tau \big)(z_1,z_3) + \mathrm{perm.}\Big\rbrace \,.
\label{eq:2pointflow}
\end{align}
It depends not only on the propagator $G_\tau$, but also on
$\Gamma^{(3)}_\tau$ and $\Gamma^{(4)}_\tau$.  We discuss the
truncation we used to solve it in section \ref{sec:scalar_field}.

The fact that higher-order correlation functions enter in the flow of the lower-order ones is a general feature of fRG equations. Typically, the flow of $\Gamma^{(n)}_{\tau}$ contains contributions from $\Gamma^{(n+1)}_{\tau}$ and $\Gamma^{(n+2)}_{\tau}$. In particular, non-Gaussian initial conditions are implemented with non-trivial $\Gamma^{(n>2)}_{\tau=t_0}$ at the initial time $t_0$.

\subsection{Causality}
\label{sec:causality}

We implicitly defined the regulator by requiring that no information
after the cutoff time $\tau$ can contribute to correlation functions
with time arguments earlier than $\tau$. This does not determine the
regulator uniquely but a simple and useful choice is given by the
sharp regulator
\begin{align}
	\imag R_{\tau}(x,y) = \delta_{\contourC}(x-y)
		\begin{cases}
			\infty 	& \text{if $x_0 = y_0 > \tau$} \\
			0 		& \text{otherwise}
		\end{cases}\,.
		\label{eq:regulator}
\end{align}
The delta distribution $\delta_{\contourC}$ on the closed time path is
defined, so that
\begin{align}
	\int_{\contourC(y)}\delta_{\contourC}(x-y)f(y) = f(x)\,.
\end{align}
A particular important consequence of causality is that the regulated
propagator $\langle \varphi(x)\varphi(y)\rangle_\textrm{1PI}$ is
strictly zero if at least one time argument exceeds the
cutoff-time. In turn, for vanishing or constant backgrounds $\phi_c$ it
is the fully interacting propagator $G\equiv G_{\tau=\infty}$
otherwise. We write the regularised propagator as
\begin{align}
    G_\tau(x, y) = G(x, y)\theta(\tau-x^0)\theta(\tau-y^0)\,. 
    \label{eq:propagator}
\end{align}
\Eq{eq:propagator} entails the important property that the full
propagator for vanishing or constant fields $\phi_c$ only depends on
the full cutoff independent two-point function $\Gamma^{(2)}[\phi_c]$
at $\tau=\infty$ with
\begin{align}\label{eq:GammanInfty}
  \Gamma^{(n)}[\phi]=\Gamma^{(n)}_{\tau=\infty} [\phi]\,.
\end{align}
Note that for general space-time dependent backgrounds, this cannot
hold true as such backgrounds can have support for all times.

The surprising property \eq{eq:propagator} is deeply rooted in the
\textit{locality} and \textit{causality} of the present cutoff
procedure. Furthermore, it is linked to the functional optimisation of
the fRG, \cite{Pawlowski:2005xe}. There, it has been shown that
optimised fRG flows have a related property: for optimal cutoffs, the
regulator variation of the two-point function perpendicular to the
direction of the optimised flow vanishes:
$\delta_\bot \Gamma^{(2)}_k =0$. The \textit{local} temporal
regularisation discussed in the present work shares this property.

A useful alternative representation of \eq{eq:propagator} is given by
\begin{align}
  G_\tau(x, y) = \left[\frac{\imag}{\Gamma^{(2)}+R_\tau}\right](x,y)\,. 
    \label{eq:propagatorAlt}
\end{align}
The causal structure extends to all correlation functions
$\langle \varphi(x_1)\cdots\varphi(x_n)\rangle_\textrm{1PI}$: they are
fully dressed as long as all their time arguments are smaller or equal
to $\tau$ and the initial ones otherwise. Hence,
$\Gamma^{(n)}_\tau[\phi_c]$ for constant backgrounds $\phi_c$ with
$n\geq2$ can be written as
\begin{align}\nonumber
  \Gamma^{(n)}_\tau(x_1,\dots,x_n) =
  &\,\Gamma^{(n)}_{t_0}(x_1,\dots,x_n) \\[1ex]
  &\,	+ \, \Delta \Gamma^{(n)}(x_1,\dots,x_n)
    \prod_{i=1}^n \theta(\tau-x^0_i) \,,
	\label{eq:causalcorrelator}
\end{align}
with
\begin{align}\label{eq:DeltaG}
  \Delta \Gamma^{(n)} = \Gamma^{(n)} - \Gamma^{(n)}_{t_0}\,. 
\end{align}
For more details see also appendix~B of \cite{Gasenzer:2010rq}. These
are the causal properties discussed in \ref{sec:closed_time_path} in
terms of 1PI correlators. They are preserved by the flow equation and
make this approach manifestly causal.

The causality constraints of the present temporal fRG also lead to
another very important identity that is peculiar to our approach: for
constant backgrounds, the line with the cutoff insertion,
$G_\tau \cdot \partial_\tau R_\tau\cdot G_\tau$ is simply given by the
$\tau$-derivative of the propagator,
\begin{align}
  \partial_\tau G_\tau(x, y) = \imag
  \int\displaylimits_{\mathrlap{\contourC(z_1,z_2)}}
  G_\tau(x,z_1)\,\partial_\tau R_\tau(z_1,z_2) G_\tau(z_2,y)\,.  
    \label{eq:regulator_derivative}
\end{align}
\Eq{eq:regulator_derivative} follows readily from the $\tau$-derivative of \eq{eq:propagator} using the representation \eqref{eq:propagatorAlt}. In contrast to standard flows with momentum cutoffs, the term proportional to $\partial_\tau \Gamma^{(2)}_\tau$ is absent. \Eq{eq:regulator_derivative} has important implications on the general structure of the temporal flow equations and is crucial for the approach.

We emphasise that the renormalisation of the theory is done for the initial state, or more precisely, for the correlation functions at the initial time. In contradistinction to time-dependent renormalisation procedures, such a time-independent one naturally incorporates the underlying RG-invariance.  
More details will be presented in \cite{CausalFlowRen,CausalFlowEMT}.

\subsection{Integrated flow}
\label{sec:integrated_flow}

The causality of the flow equation, in particular, the properties of
the propagator \eqref{eq:propagator} and the higher 1PI correlation
functions \eqref{eq:causalcorrelator}, as well as the relation for the
regulator derivative \eqref{eq:regulator_derivative} have the
remarkable consequence that the time flow can always be integrated
analytically.

\begin{figure}[t]
	\centering
	\includegraphics[width=0.45\textwidth]{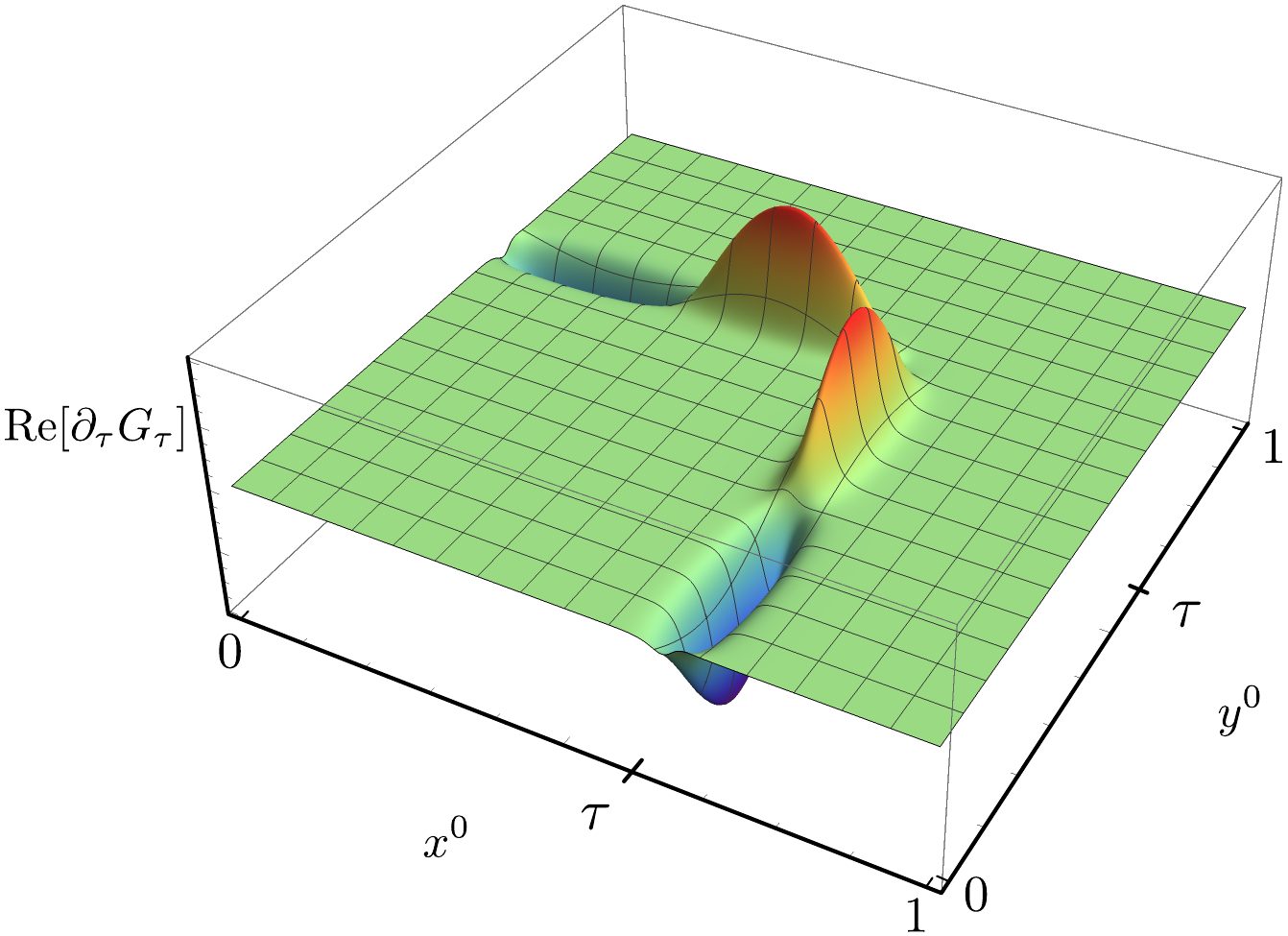}
	\caption{Illustration of the causal structure of the
          propagator derivative $\partial_\tau G_\tau$
          (cf.~\eqref{eq:propagator_derivative}). For better
          visualisation, we introduced a finite width to the
          contributing $\theta$- and $\delta$-functions. The
          oscillating structure is the real part of the free
          propagator (both field operators are inserted on the forward
          branch of the CTP but other insertions give a similar
          picture). The imaginary part basically only differs by a
          phase. }
	\label{fig:causality_illustration}
\end{figure}

This integration is possible since all flow equations for the
$\Gamma^{(n)}_\tau$ can only contain the regulator derivative in the
form of \eqref{eq:regulator_derivative} which can be replaced by the
$\tau$-derivative of the propagator
\begin{align}
  \nonumber
  \partial_\tau G_\tau(x, y)
  = G(x, y) \Big[
  &\delta(\tau-x^0)\theta(\tau-y^0) \\[1ex]
  &+\theta(\tau-x^0)\delta(\tau-y^0)\Big]\,.	
	\label{eq:propagator_derivative}
\end{align}
A visualisation of \eq{eq:propagator_derivative} is depicted in
\fig{fig:causality_illustration}. For illustration purposes, we
introduced a finite width for the $\theta$- and $\delta$-functions and
plugged in the free propagator for $G$. In
\fig{fig:causality_illustration}, we show the real part of the free
propagator (both field operators are inserted on the forward branch of
the CTP but other insertions give a similar picture). The imaginary
part basically only differs by a phase.
 
Now, consider the first line of \eqref{eq:propagator_derivative}. At
$x^0 = \tau$, the propagator derivative in $y^0$-direction carries the
oscillating shape of the real part of the free propagator. However, as
soon as $y^0 > \tau$, the propagator vanishes. The same holds if we
swap $x^0$ and $y^0$.

Taking a $\tau$-derivative of \eqref{eq:causalcorrelator}, we obtain
relations similar to \eqref{eq:propagator_derivative} for the
$\Gamma^{(n)}_\tau$. Therefore, the full $\tau$\nobreakdash-dependence
of the flow equations is captured by $\theta$- and
$\delta$\nobreakdash-distributions. This guarantees that the
$\tau$-integration can always be performed analytically.

Strikingly, the resulting equations are one-loop equations for the
fully dressed correlation functions. This is particularly useful when
the theory has to be renormalised as the divergent contributions are
readily identified.  Further note that this is really due to
causality. For example, a sharp cutoff regulator in momentum space
does not lead to an integrated flow that is one-loop.

We demonstrate the analytic integration for the case of the two-point
function in \app{app:t-fRGPropagator}.

\section{Dynamics of the \texorpdfstring{$\phi^3$}{phi\textasciicircum
    3}-theory}
\label{sec:scalar_field}

In this section we implement the t-fRG approach for a scalar field
with cubic interaction in $1+1$ dimensions with the classical action
\begin{align}
  S[\varphi]=\int_{\contourC(x)}\bigg\lbrace\frac{1}{2}
  \partial^\mu\varphi(x)\partial_\mu\varphi(x)-
  \frac{m^2}{2}\varphi(x)^2-\frac{\lambda}{3!}\varphi(x)^3\bigg\rbrace\,.
    \label{eq:action}
\end{align}
As discussed in \sec{sec:introduction}, this theory is an ideal test
case for the present approach. Moreover, microscopic cubic
interactions are also present in non-Abelian gauge theories. Even
though the latter are momentum dependent, the scalar field theory with
\eq{eq:action} allows for the same scattering processes. Since those
scattering processes are absent in the $\phi^4$-theory, the insights from
cubic interactions are a necessity in regard to non-Abelian gauge
theories.

\subsection{Truncated flow for the propagator}
\label{sec:truncatedflow}
\begin{figure}[t]
	\centering
	\includegraphics{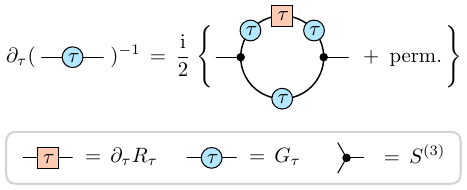}
	\caption{Truncated flow equation for the two-point function. The
          orange square denotes $\partial_\tau R_\tau(x,y)$.  Lines
          with blue circles represent the fully dressed propagator
          ${G_\tau(x,y)~=~\imag\big[\Gamma^{(2)}_\tau+
            R_\tau\big]^{-1}(x,y)}$. The black dot with three legs is
          the classical vertex $S^{(3)}$. Since we further put
          $\Gamma^{{(n)}}_\tau = 0$ for all $n>3$, all other
          diagrams vanish. perm. stands for the permutation of the
          external legs.}
	\label{fig:flow_equation_truncated}
\end{figure}

In the present work we discuss the dynamics of the propagator with
classical three-point functions. In contradistinction to the
$\phi^4$-theory, in the $\phi^3$-theory this already gives rise to a
non-trivial dynamical evolution. 
The vertex reads
\begin{align}
	\nonumber
	\Gamma^{{(3)}}_\tau(x,z_{1},z_{2})
		&= S^{{(3)}}(x_,z_{1},z_{2}) \\[1ex]
		&= -\lambda \deltac(x-z_{1})\deltac(z_{1}-z_{2})\,.
\label{eq:three_point_function}
\end{align}
Additionally, we set $\Gamma^{{(n)}}_\tau = 0$ for all
$n>3$. This leads us to the truncated flow equation for the two-point
function shown in \fig{fig:flow_equation_truncated}.
In this approximation, we study the emergence of universal dynamics in the $\phi^3$-theory. 
This universal dynamics can be generated with Gaussian initial conditions (see previous 2PI--studies of $\phi^4$-theory, e.g. \cite{Berges2017, Shen:2019jhl}).

Let us emphasise again that including non-Gaussian initial conditions is straightforward in the t-fRG framework in the form of $\Gamma^{(n>2)}_{\tau=t_0}$. The respective higher order correlations are e.g. part of a thermal state given by a density matrix of the form ${\rho\propto\exp^{-\beta H}}$ with the Hamiltonian $H$. Hence, the Gaussian initial state maybe understood as the result of some type of quench which drives the system out of equilibrium.

If we further insert \eqref{eq:regulator_derivative} for the part
including the regulator derivative, we obtain
\begin{align}
\partial_\tau\Gamma^{(2)}_\tau(x,y)
	&= \frac{\imag\lambda^2}{2}
		\partial_{\tau}G_{\tau}^{2}(x,y)\,.
		\label{eq:flow_gamma2}
\end{align}
Integrating over $\tau$ is now straightforward. Due to our simple
truncation, the flow \eqref{eq:flow_gamma2} is in fact a total
$\tau$\nobreakdash-derivative. Let us stress again that the flow can also be
integrated analytically in general truncations.  This is discussed in
detail in \App{app:t-fRGPropagator} and is a consequence of
the causal structure of the flow. Performing the integration yields
\begin{align}
  \Gamma^{(2)}(x,y) = \Gamma^{(2)}_{t_0}(x,y) +
  \frac{\imag\lambda^2}{2}G^{2}(x,y) \,,
	\label{eq:integrated_flow}
\end{align}
where $\Gamma^{{(2)}}_{t_0}$ is the free kinetic operator of
\eqref{eq:action}. Thus, \eqref{eq:integrated_flow} can be solved as
an integro-differential equation by applying it to the propagator
$G$. This approach is very common in the literature. In our case, it
yields the following equation
\begin{align}\nonumber 
  \left[\,\partial_{x}^2 + m^2\,\right]G(x,y)
  =&\, -\imag \delta_{\contourC}(x - y) \\[1ex] 
 & + \frac{\imag\lambda^2}{2} \!
  \int\displaylimits_{\mathrlap{\contourC(z)}}
  G^2(x,z)G(z,y) \,.
	\label{eq:diffeq}
\end{align}
Let us remark that equation \eq{eq:diffeq} can also be obtained as the lowest-order in 2PI--perturbation theory. The t-fRG framework can reproduce 2PI--approximations by a suitable choice of truncation, cf. the discussion in \cite{Gasenzer:2007za, Gasenzer:2010rq} regarding the s-channel resummation.
In general, however, the t-fRG framework allows for resummations that do not correspond to 2PI resummations, e.g. by including t- and u-channel contributions of the four\nobreakdash-point function.
These options will be explored in future work.

Another possibility is to invert the free kinetic operator
$\Gamma^{(2)}_{t_0}$ using
\begin{align}
  \int\displaylimits_{\mathrlap{\contourC(z)}}
  \Gamma^{(2)}_{t_0}(x,z)\bar{G}(z,y) =
  \imag\delta_\contourC(x-y)\,,
\end{align}
where we denoted the inverse by $\bar{G}$. Since $\bar{G}$ is just the
solution of the free equation of motion, it is known
analytically. Multiplying \eqref{eq:integrated_flow} with $\bar{G}$
from the left and $G$ from the right, we obtain the integral equation
\begin{align}
  G(x,y) =\bar{G}(x,y) - \frac{\lambda^2}{2}
  \int\displaylimits_{\mathrlap{\contourC(z_1,z_2)}} \bar{G}(x,z_{1})
  G^2(z_{1},z_{2})G(z_{2},y)\,.
\label{eq:dysoneq}
\end{align}
We also remark that while \eq{eq:dysoneq} seemingly is an
implicit equation, it is in fact explicit for $x^{0}\geq y^{0}$. 
Using the symmetry of the propagator, this
allows us to solve \eq{eq:dysoneq} explicitly without iterating it.

Let us briefly address the numerical solution of these equations. At a
first glance, the integro-differential version of the equation can be
solved faster since there is one time integral less compared to the
integral version. However, demanding that the results have the same
accuracy, this changes. Due to the derivative, a higher resolution is
needed to achieve the same accuracy as with the integral
equation. More details can be found in appendix \ref{app:numerics}.

\subsection{Results}
\label{sec:results}
\begin{figure}[t]
  \centering \includegraphics{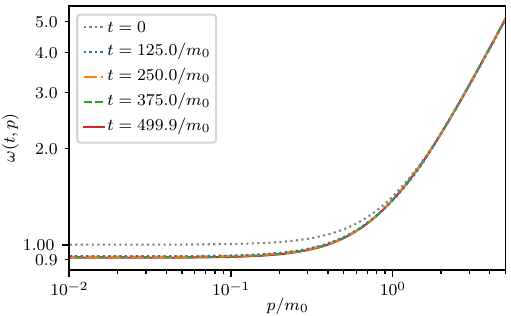}
  \caption{Time evolution of the dispersion relation as defined in
    \eqref{eq:omega}. It is depicted by showing the dispersion
    relation for various times. The grey dashed line corresponds to
    the dispersion at the initial time.  At zero-momentum one can read
    off the mass of the interacting theory to be $m\approx 0.9m_{0}$
    relative to the bare mass.}
	\label{fig:dispersion}
\end{figure}
For the solution of the integrated flow, it is useful to formulate all
equations in terms of the statistical two-point function $F$ and the
spectral function $\rho$. These are defined as the expectation value
of the anti-commutator and commutator of the field, respectively and 
(for $\phi = 0$) read
\begin{align}\nonumber 
  F(x,y)
  &= \frac{1}{2}\big\langle\lbrace {\varphi}(x),{
    \varphi}(y)\rbrace\rangle\,, \\[1ex]
  \rho(x,y)
  &= \imag\big\langle[{\varphi}(x),
    {\varphi}(y)]\big\rangle\,.
\end{align}
From this definition, it is clear that $F$ is symmetric, while $\rho$
is anti-symmetric. They are related to the propagator by
\begin{align}
	G(x,y) = F(x,y) - \frac{\imag}{2}\rho(x,y)\sgnc(x^0,y^0)\,,
\end{align}
where $\sgnc(x_0-y^0)$ is $1$ for $x^0> y^0$ on the CTP and $-1$ for
$x^0< y^0$.

The propagator already allows us to discuss relevant observables such
as the occupation number and the dispersion relation. In
non-equilibrium situations, there is no unique definition, but we can
define versions analogously to the ones in equilibrium. To that end, we
make use of the decomposition of the
equal time statistical propagator, see e.g.~\cite{Berges2005}, 
\begin{align}
	F(t,t,\vec{p}) = \frac{f(t,\vec{p}) + \frac12}{\omega(t,\vec{p})}\,.
\end{align}
The non-equilibrium generalisations of the occupation number
$f(t,\vec{p})$ and dispersion relation $\omega(t,\vec{p})$ are chosen
such that they coincide with their time-independent counterparts in
equilibrium. The occupation number can be computed as
\begin{align}
	f(t,\vec{p})
	= \big[\partial_{t}\partial_{t'}F(t,t';\vec{p})\vert_{t'=t}F(t,t;\vec{p}) 
	\big]^{1/2} - \frac{1}{2} \,,
	\label{eq:f}
\end{align}
and for the dispersion relation one finds
\begin{align}
  \omega(t,\vec{p}) = \left(\frac{\partial_{t}
  \partial_{t'}F(t,t';\vec{p})\vert_{t'=t}}{F(t,t;\vec{p})}\right)^{1/2} \,.
	\label{eq:omega}
\end{align}
In \app{app:init} we exemplify the above definitions with the
solution of the free equation of motion. There, one can directly verify
that the definitions \eqref{eq:f} and \eqref{eq:omega} give the
desired results.

For the results shown in this section, we choose initial conditions
far from equilibrium: we prepare a system with highly over-occupied
momentum modes at small momenta and none for high momenta.
Explicitly, we consider a (sharp) box for the initial occupancies
$f_{0}(\vec{p})$ of the form
\begin{align}
  f_{0}(\vec{p}) = \frac{\widetilde{N}}{\widetilde{\lambda}}
  \theta(m_{0}-\vert\vec{p}\vert)\,.
\end{align}
Here, $\widetilde{N}=100$ and the dimensionless coupling of the three-point
function is given by $\widetilde{\lambda}=\lambda/m_0^2=0.01$. More
details on the initial conditions for the different solvers are given
in \app{app:init}. The results of this section were obtained
solving the integro-differential version of the equation.

A first interesting result is the time evolution of the dispersion
relation shown in \fig{fig:dispersion} for different times.  At
small momenta the dispersion decreases with time. This region is
dominated by the mass. For zero momentum we can therefore read off the
mass $m$ of the interacting particles compared to the bare mass $m_0$
and find ${m \approx 0.9m_0}$. For higher momenta, where the mass is
negligible, the dispersion agrees for all times.

\begin{figure}[t]
	\centering
	\includegraphics{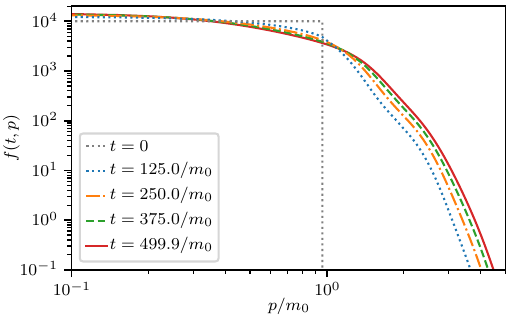}
	\caption{Time evolution of the occupation number as defined in
          \eqref{eq:f}. It is represented by showing the occupation
          number for various times. The grey dashed line corresponds to the
          initial occupations. For later times the initial box is smoothed
          out and we find indications for a self-similar scaling regime around
          ${p/m_{0}\approx 2}$ exhibiting a power law decay.}
	\label{fig:occupation_number}
\end{figure}

The time evolution of the occupation number is shown in
\fig{fig:occupation_number} for the same times as used for the
dispersion relation. Naturally, the initial sharp box is softened
during the time evolution, and particles are redistributed over the
range of momenta.

The momentum regime around $p/m_{0}\approx 2$ is particularly
interesting. In this regime we may identify self-similar scaling with a
power law decay of the occupation number
\begin{align}
	f(t,\vec{p}) \propto \vert\vec{p}\vert^{-\kappa} \,.
	\label{eq:powerlaw}
\end{align}

\begin{figure}[t]
	\centering
	\includegraphics{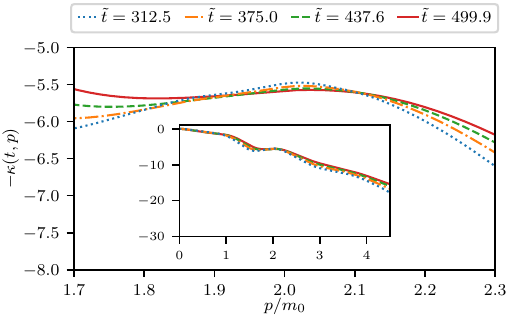}
	\caption{Momentum-dependent exponents of the occupation number
          for various times $\tilde{t}=tm_{0}$ as defined in
          \eqref{eq:momentum_dependent_exponents}. In the momentum
          range $p/m_0 \in [1.8,2.1]$, the exponents are approximately
          constant. At later times this constant regime is more
          pronounced, and we have evaluated the exponent at
          $t=499.9/m_0$ as $\kappa\in [5.57\,,\,5.69]$. The inset
          shows the momentum-dependent exponents for the full
          available momentum range.}
	\label{fig:exponents}
\end{figure}

For an estimate of the exponent $\kappa$, we compute the
momentum-dependent exponent
\begin{align}
  \kappa (t, p) = -p \,\partial_{p}\ln\, f(t,p)\,.
	\label{eq:momentum_dependent_exponents}
\end{align}
This exponent is shown for different times in \fig{fig:exponents}. In
the momentum range $p/m_0 \in [1.8,2.1]$ this exponent is
approximately constant. At later times this constant scaling regime is
more pronounced, and we have evaluated the exponent at $t=499.9/m_0$
for the momentum range above as
\begin{align}\label{eq:kappa}
  \kappa\in [5.57\,,\,5.69]\,, \qquad \frac{p}{m_0} \in [1.8\,,\,2.1]\,.
\end{align}
The analysis above suggests a power law behaviour.  Moreover, the
exponent is similar for all times considered. This indicates a
self-similar scaling, although the regime is rather small.

\begin{figure}[t]
  \centering \includegraphics{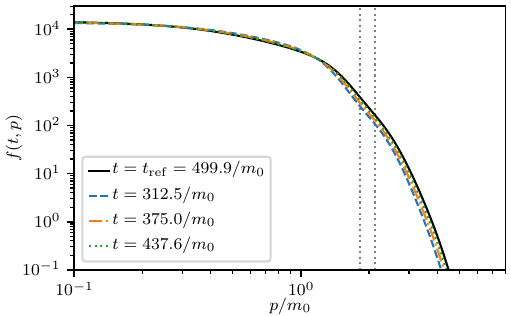}
  \caption{Occupation numbers for times for which a
      self-similar time evolution is indicated in the momentum range
      marked by vertical dashed lines.}
	\label{fig:original_occupation_number}
\end{figure}

\begin{figure}[t]
	\centering
	\includegraphics{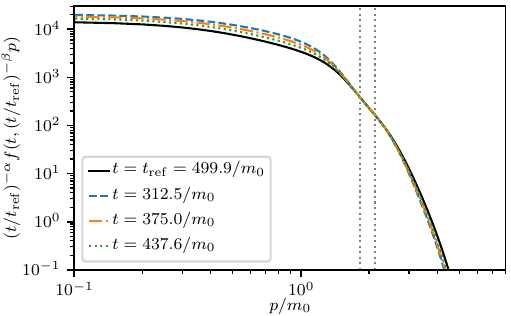}
	\caption{Occupation numbers rescaled according to
          \eq{eq:selfsimilar} for an assumed self-similar time
          evolution in the momentum range $p/m_{0}\in[1.8,2.1]$. This
          regime is marked by vertical dashed lines. The $\alpha,
          \beta$-exponents computed for times $tm_{0}=$312.5,\,375.0,\,437.6
          are given by $\alpha=$0.82,\,1.03,\,1.39 and
          $\beta=$$-0.02$,\,0.02,\,0.09.}
	\label{fig:rescaled_occupation_number}
\end{figure}

In regimes with self-similar scaling, the time evolution is
characterised by a self-similar scaling of the occupancies, see
eg.~\cite{PineiroOrioli2015,Schmied:2018mte}. This scaling reads
\begin{align}
  f(t_{\mathrm{ref}},\vert\vec{p}\vert) = \left(\frac{t}{
  t_{\mathrm{ref}}}\right)^{-\alpha} f\left[t,\bigg(\frac{t}{
  t_{\mathrm{ref}}}\bigg)^{-\beta} \vert\vec{p}\vert\right] \,.
  \label{eq:selfsimilar}
\end{align}
In the regime $p/m_0 \in [1.8,2.1]$, we find for the times
${tm_{0}=312.5,\,375.0,\,437.6}$ the exponents
${\alpha=0.82,\,1.03,\,1.39}$ and ${\beta=-0.02,\,0.02,\,0.09}$,
employing a least squares fit with respect to the occupancies at the
reference time $t_{\mathrm{ref}} = 499.9/m_0$.
\Fig{fig:original_occupation_number} shows the original occupation
numbers for the above times and \fig{fig:rescaled_occupation_number}
the ones rescaled each by their corresponding exponents.  The rescaled
occupation numbers match in the momentum range found from the power
law exponent in accordance with a self-similar time evolution.

Next, we discuss the non-trivial and important consistency check of 
energy conservation for the present computation. The total
energy is obtained by computing the expectation value of the time-time
component of the energy-momentum tensor $T_{\mu\nu}$.  The details of
this computation are discussed in \app{app:T00}, where a
representation of the energy solely in terms of the propagator is
derived, see \eq{eq:EMT-Exp-Reduced}. Using $\rho(t,t,\vec{p}) = 0$ in
\eqref{eq:EMT-Exp-Reduced}, we find for the total energy
\begin{align}\nonumber 
  E(t) = \langle T_{00}(t)\rangle =
  &\, 
    \lim_{t\to t'} \frac56   \partial_{t} \partial_{t'} \int_{\vec p}
    F(t,t';{\vec p})\\[1ex]
  &+ \frac16   \int_{\vec p}\left( {\vec p}^2 +m_0^2\right) 
    F(t,t;{\vec p})\,.
   \label{eq:EMT-ExpF}
\end{align}

The explicit occurrence of $\Gamma^{(3)}$ has dropped out in
\eq{eq:EMT-ExpF} thanks to the structure of the gap equation (Dyson-Schwinger equation) for the two\nobreakdash-point function \eq{eq:DSE} in $\phi^3$\nobreakdash-theory.

\begin{figure}[t]
	\centering
	\includegraphics{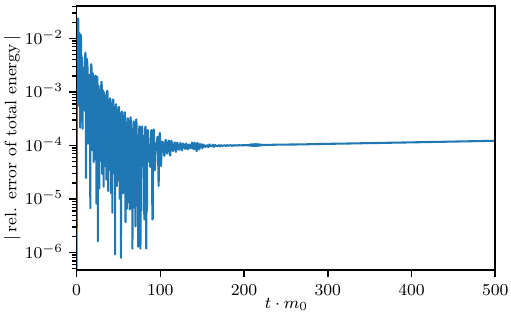}
	\caption{Relative error of the total energy with respect to
          the initial total energy
          $\big\vert \big[E(t)-E(0)\big]/E(0)\big\vert$. After an
          initial tune in, the relative error stabilises at around
          $10^{-4}$.}
	\label{fig:relative_error_energy}
\end{figure}

As was already mentioned, the truncation employed in the present work corresponds to the lowest-order in 2PI--perturbation theory. The latter is known to maintain energy conservation. However, this may be violated in numerical implementations. As a consistency check, we show the relative error of the total energy in \fig{fig:relative_error_energy}. The error stabilises
at around $10^{-4}$ and the total energy is conserved.

While an analytical proof of energy conservation for general t-fRG truncations is lacking, the causal and local nature of the temporal flow provide a handle to address the question of analytically identifying conserving truncations. This is ongoing work \cite{CausalFlowEMT}.

\begin{figure}[t]
	\centering
	\includegraphics{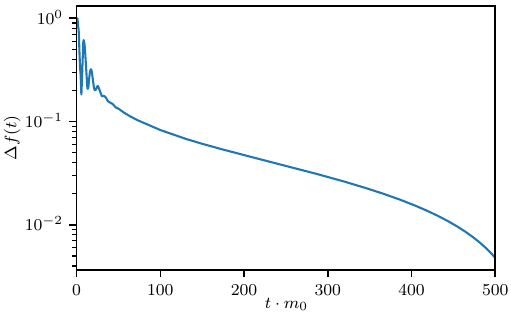}
	\caption{Sum of positive and negative flows in particle
            number normalised to the difference of positive and
            negative flow (total flow) of particle number as defined
            in \eqref{eq:particle_number_error}. After initial
            oscillations the total particle number is conserved.}
	\label{fig:particle_number_conservation}
\end{figure}

For the discussion of particle number conservation we use
\begin{align}
	\Delta f(t) = \frac{\int_p f(t,p) - f(0,p)}{\int_p \vert f(t,p) - f(0,p) \vert} \,,
	\label{eq:particle_number_error}
\end{align}
which measures the sum of positive and negative flow of particle
numbers normalised to the difference of positive and negative flow of
particle numbers (total flow). \Fig{fig:particle_number_conservation}
shows this quantity and we find, similarly to the total energy, that
after the initial oscillations, the total particle number is
conserved.

\section{Conclusion}  
\label{sec:conclusion}
In the present work, we have further developed the temporal functional
RG (t-fRG) suggested in \cite{Gasenzer:2007za, Gasenzer:2010rq}. By
integrating the flow, we have derived novel one-loop exact functional
relations for correlation functions. This result originates in the
causal structure of the flow and its locality in time.

Moreover, we have applied the approach in its integrated form to the
$\phi^3$-theory in 1+1 dimensions.  This theory serves as a test case
for the approach and simulates the cubic interactions in a non-Abelian
gauge theory. Here, we have studied the simplest approximation of the
dynamics of the propagator, using classical vertices in the integrated
flow. This approximation leads to a dynamical resummation of the
propagator. We find indications of a scaling regime out of equilibrium
with a momentum scaling $|{\vec p}|^{-\kappa}$ with
$\kappa\in [5.57\,,\,5.69]$ at $t=499.9/m_0$, see \eq{eq:kappa}. An
investigation of the self-similarity of the three times available led
to the coefficients $\alpha\in [0.82\,,\,1.39]$ and
$\beta \in [-0.02\,,\, 0.09]$, see \eqref{eq:selfsimilar}. Further
investigations of this regime as well as the extension to situations
with wave-turbulence are deferred to future work.

There are several interesting extensions of the present work. The
present framework allows for rather general -causal- approximation
schemes, most notably the extension to non-trivial vertices and
non-vanishing backgrounds. This is work in progress and we hope to
report on the results in the near future.

\acknowledgments

We thank J\"urgen Berges, Thomas Gasenzer, Eduardo Grossi, Linda Shen
and Nicolas Wink for discussions. This work is supported by EMMI, the
BMBF grant 05P18VHFCA, and is part of and supported by the DFG
Collaborative Research Centre SFB 1225 (ISOQUANT) as well as by the
DFG under Germany's Excellence Strategy EXC - 2181/1 - 390900948 (the
Heidelberg Excellence Cluster STRUCTURES).

\appendix

\section{Integrated flow for the propagator}
\label{app:t-fRGPropagator}

Here, we discuss the analytic integration of the general flow equation
of the two-point function with the help of causal
relations. Remarkably, we obtain a one-loop exact functional relation
valid for general theories. The derivation makes no use of the
interaction structure of the theory. In particular, it applies to
$\phi^3$- and $\phi^4$-theories used as explicit -and general-
examples in this section.

We start with the full flow of the two-point function
\eqref{eq:2pointflow}.  With \eqref{eq:regulator_derivative} we
replace $G_\tau\partial_\tau R_\tau G_\tau$ by
$-\imag\partial_\tau G_\tau$and are led to 
\begin{align}\nonumber
  \partial_\tau \Gamma^{(2)}_\tau(x,y) =  
  & \,\frac12 \int\displaylimits_{\mathrlap{\contourC(z_1,z_2)}}
    \Gamma^{(4)}_\tau (x,y,z_1,z_2) 
    \partial_\tau G_\tau (z_2,z_1) \\[1ex]	 \nonumber
  + &\, \frac\imag2 \int\displaylimits_{\mathrlap{\contourC(z_1,\dots,z_4)}}
      \hspace{1.2cm}
      \Biggl[ \Gamma^{(3)}_\tau(x,z_1,z_2) G_\tau(z_2,z_4) \big.  \\[1ex]
  &\,\hspace{-.8cm} \times \, \Gamma^{(3)}_\tau(z_3,z_4,y)
    \partial_\tau G_\tau(z_1,z_3)
    + \mathrm{perm.}\Biggr] \,.
	\label{eq:twopointflow_Gdot}
\end{align}
From now on, we use a condensed notation with space-time arguments as
indices.  This makes the structural aspects of the following arguments
more apparent. The first term on the right hand side of
\eqref{eq:twopointflow_Gdot} can be rewritten as a total
$\tau$-derivative and a $\partial_\tau\Gamma^{(4)}_\tau$-term,
\begin{align}\nonumber
  \frac12 \int\displaylimits_{\mathrlap{\contourC(z_1,z_2)}}
  \Gamma^{(4)}_{\tau, xyz_1z_2} 
  \partial_\tau G_{\tau, z_2z_1} 
  = & \partial_\tau\left[\frac12\int\displaylimits_{\mathrlap{\contourC(z_1,z_2)}} 
      \Gamma^{(4)}_{\tau, xyz_1z_2}  G_{\tau, z_2z_1}\right]\\[1ex]
  - & \frac12 \int\displaylimits_{\mathrlap{\contourC(z_1,z_2)}} 
      \partial_\tau\Gamma^{(4)}_{\tau, xyz_1z_2}  G_{\tau, z_2z_1} \,.
	\label{eq:tadpole}
\end{align}
\Eq{eq:tadpole} relates to the fact that the flow equation for the
effective action can be rewritten as a total $\tau$-derivative and an
RG improvement term,
\begin{align}\label{eq:MasterFlowTotal}
  \partial_\tau \Gamma_\tau = \frac{\imag}{2} \Tr\, \partial_\tau\!
  \log \left[\Gamma^{(2)}_\tau+ R_\tau\right] -\frac{1}{2}
  \Tr \,\partial_\tau \Gamma_\tau^{(2)} G_\tau \,, 
\end{align}
where we have suppressed the field dependence. While such a rewriting
is not required for applications, it carries much of the structure of
the flow equation: typically, the first term is dominant while the
second term generates sub-leading RG-improvements.

The first term in \eq{eq:tadpole} or \eq{eq:MasterFlowTotal} can be
integrated directly. For the second term, we use the causal properties
of our approach and, in particular, \eq{eq:propagator} and
\eq{eq:causalcorrelator}.  For $n=2$, we rewrite the causal property
as
\begin{align}\nonumber 
  \partial_\tau \Gamma^{(2)}_{\tau, xy} =
  &\, \Delta \Gamma^{(2)}_{xy} * 
    \Big[\delta(\tau-x^0)\theta(\tau-y^0) \\[1ex]
  &\,\hspace{1.5cm}+ \theta(\tau-x^0)
    \delta(\tau-y^0)\Big]\,,
	\label{eq:causalgamma2dot}
\end{align}
with $\Delta \Gamma^{(2)}= \Gamma^{(2)}-\Gamma^{(2)}_{t_0}$, see
\eq{eq:DeltaG}. Here, $\Gamma_{t_0}$ is the input effective action. In
the present example, $\Gamma^{(n)}_{t_0}$ are the $n$th derivatives of
the classical action. The notation ${{\cal O}_1(x,y) * {\cal O}_2(x,y)}$
takes account of the intricacy that
$ \partial_\tau \Gamma^{(2)}_{\tau, xy} $ may contain terms such as
\begin{align}\label{eq:PreSalm}
  \lim_{r\to 0}\theta_r(\tau - x^0)\delta_r(\tau- x^0) =
  \frac12 \delta(\tau- x^0)\,, 
\end{align}
where the subscript $r$ indicates a general regularisation of the
$\theta$- and $\delta$-function.  Terms as in the example
\eq{eq:PreSalm} arise from local contributions in $\Delta\Gamma^{(2)}$
proportional to ${\deltac(x^0-z_i^0)}$ with $i=1,2$. We emphasise that
the product in \eq{eq:PreSalm} is uniquely defined as both, the
$\theta$- and the $\delta$\nobreakdash-function, share the same
regularisation as distributions, which originates in a regularisation
of the sharp temporal regulator ${R_\tau \to R_{\tau,r}}$.

This structure extends to
general $n$-point functions and we write
\begin{align}\label{eq:local}
  \Delta \Gamma^{(n)}_{x_1\cdots x_n} =
  \Delta \Gamma^{(n)}_{\mathrm{nl},x_1\cdots x_n} +
  \Delta \Gamma^{(n)}_{\mathrm{local},x_1\cdots x_n}\,,
\end{align}
where we collect all terms that include $\deltac(x_i^0-x_j^0)$ with $i\neq j$ in the
local part. We remark that for the non-local part, the
$*$\nobreakdash-product reduces to the standard product as no product
of singular distributions is present.

Note that \eq{eq:causalgamma2dot} is proportional to
$\delta(\tau-x^0)$ and ${\delta(\tau-y^0)}$. In the diagrams in
\eq{eq:causalcorrelator}, these are the external time arguments.
Using \eqref{eq:causalcorrelator} for $n=4$ and comparing with
\eqref{eq:causalgamma2dot}, we conclude that there are two types of
non-vanishing contributions of the second term of \eqref{eq:tadpole}: 

\begin{itemize}
\item[(i)] \textit{Contributions from the non-local part of the
    vertex:} $\Delta\Gamma^{(4)}_{\textrm{nl}}$ can only contribute
  if the $\tau$-derivative hits a $\theta$-function with an external
  time argument. If it hits an internal argument, the locality
  constraint of the flow is not satisfied and this contribution
  vanishes. This leads us to
\begin{align}\nonumber
  \hspace{.7cm} \Delta\Gamma^{(4)}_{\textrm{nl},xyz_1z_2}
  & \left[\delta(\tau-x^0)\theta(\tau-y^0) + x^0
    \leftrightarrow y^0\right] \\[1ex]
  \times \, & \theta(\tau-z_1^0)\theta(\tau-z_2^0) \,,
\label{eq:G4nl-Contributions}\end{align}
that carry the $\delta$-functions in \eq{eq:causalgamma2dot}. The
product in \eq{eq:G4nl-Contributions} is the standard one.

\item[(ii)] \textit{Contributions from the local part of the vertex:}
  $\Delta\Gamma^{(4)}_{\textrm{local}}$-contributions have to be
  computed with care. As for the first type in
  \eq{eq:G4nl-Contributions}, we receive contributions from
  $\partial_\tau \theta(\tau- x^0_i)$. Since additional factors
  $\theta(\tau-x^0_i)$ may be present, the $*$-product has to be
  evaluated ($i=1,2$ and $x^0_1=x^0,\, x^0_2=y^0$).
  
  Moreover, also contributions from the $\tau$-derivative of
  $\theta$-functions with an internal time argument $z_j^0$ have to be
  considered, as $\Delta\Gamma^{(4)}_{\textrm{local}}$ carries
  contributions with $\delta$-functions ${\deltac(x^0_i-z_j^0)}$ with
  $j=1,2$.  These terms also carry the required causal
  structure. Again, the $*$-product has to be considered due to the
  potential occurrence of additional $\theta$\nobreakdash-functions.
\end{itemize}

\subsection{Gap equation in the
  \texorpdfstring{$\phi^3$}{phi\textasciicircum
    3}-theory}\label{sec:gap3}

We now apply the above arguments to the two-point function of the
$\phi^3$-theory used in the present work. Integrating the flow should
result in the gap equation of the theory. While this is simply a
consistency check of the approach, the locality and causality of the
present framework will simplify the computation. We emphasise that
these simplifications are also present in numerical applications.

To begin with, the $\phi^3$-theory has the important property
\begin{align}\label{eq:NonLocal0}
  \Delta\Gamma^{(n)}_{\textrm{local}}\equiv 0\,,\quad \forall \, n>2\,. 
\end{align}
\Eq{eq:NonLocal0} greatly simplifies the following derivations and
allows us to concentrate on some important aspects. For the -sketchy-
proof of \eq{eq:NonLocal0} in the $\phi^3$-theory, we first consider
the simplest vertex corrections for the three-point function, the
triangle diagram with classical vertices. Thus, the vertex correction
is given by 
\begin{align}\label{eq:triangle} 
  \Delta\Gamma^{(3)}_{x_1 x_2 x_3} \propto
  \lambda^3 G(x_1,x_2) G(x_1,x_3) G(x_2,x_3)+ \cdots \,, 
\end{align}
where the dots stand for other diagrams as well as vertex corrections
in the triangle. The space-time dependence of \eq{eq:triangle} is
simply given by a product of propagators. Evidently, this product does
not contain temporal or spatial $\delta$-functions, as long as it is
well-defined. 

We add that in dimensions $d\geq 1+5$, the product of propagators in
\eq{eq:triangle} is not well-defined anymore at $x_1=x_2=x_3$. The
respective terms are proportional to $\deltac(x_1-x_2)\deltac(x_2-x_3)$
and hence add to the classical coupling. This is nothing but the
standard renormalisation. A more detailed account of this will be
presented elsewhere. In the present work, we consider $d=1+1$ and
these intricacies are absent.

In summary, the vertex correction \eq{eq:triangle} has no local
pieces. Furthermore, any other vertex correction to the three-point
function can be iteratively constructed from this diagram and the
respective ones for the higher correlation functions. None of these
diagrams can generate $\delta$\nobreakdash-functions, as all legs are
connected by propagators. In conclusion, the $n$-point functions in a
$\phi^3$-theory have no local parts, except the initial (classical)
vertex $\Gamma^{(3)}_{t_0}$.

This greatly simplifies the current investigation, as the local
contributions are absent and no $*$-product has to be
considered. Moreover, utilising the manifest causality of the present
approach will be pivotal for a simple derivation of the the final
result. Without loss of generality, we consider
$\Gamma_\tau^{(2)}(x,y)$ for $x^0>y^0>t_0$. The full $\Gamma^{(2)}$ is
obtained from
\begin{align}\label{eq:FlowG2}
  \Gamma^{(2)}(x,y)=\Gamma^{(2)}_{t_0}(x,y)+
  \lim_{\epsilon\to 0_+} \int\displaylimits_{t_0}^{x^0+\epsilon} \mathrm{d}\tau\,
  \partial_\tau \Gamma_\tau^{(2)}(x,y)\,, 
\end{align}
The upper boundary in \eq{eq:FlowG2} follows from the causality of the
flow leading to $\partial_\tau \Gamma^{(2)}_{\tau}(x,y) \equiv 0$ for
${\tau>x^0>y^0}$. Due to causality, we could also move the
lower boundary from $t_0\to x^0-\epsilon$.
The infinitesimal shift with $\epsilon$ has been
introduced as the flow has a contribution proportional to $\delta(\tau -x^0)$.
Indeed, due to causality and locality the flow is only non-vanishing
for $\tau=x^0$. For $y^0 > x^0$ the flows is only non-vanishing for
$\tau = y^0$. Inserting the right-hand side of \eqref{eq:tadpole} in
\eq{eq:FlowG2} leads us to a vanishing tadpole contribution,
\begin{align}\nonumber 
  &	\frac12\int_{\contourC(z_1,z_2)\leq x^0}  \bigg[
    \Gamma^{(4)}_{xyz_1z_2}  G_{z_1 z_2} 
   \\[1ex] 
  &\hspace{1cm} - 
    \Gamma^{(4)}_{t_0, xyz_1z_2} G_{t_0,z_1 z_2} 	- 
    \Delta\Gamma^{(4)}_{xyz_1z_2}  G_{z_1 z_2}\bigg] =  0 \,. 
\label{eq:tadpole0}\end{align}
\Eq{eq:tadpole0} follows from $ \Gamma^{(4)}_{t_0}=0$ in
the $\phi^3$-theory: Accordingly,
$ \Delta \Gamma^{(4)} = \Gamma^{(4)}$ and the first and third term
cancel. The second term vanishes trivially: both $\Gamma_{t_0}^{(4)}$
and $G_{t_0}$ vanish separately.

We emphasise that the property \eq{eq:tadpole0} is unique to the
present t-fRG approach and is a consequence of both locality and
causality. For other regulator choices and, in particular, the common
momentum and frequency regulators, the tadpole contribution is
non-vanishing. To obtain the gap equation from the integrated combined
flow in such a setting, one has to insert the Dyson-Schwinger equation
for the four-point function in the tadpole and proceed from
there. Naturally, \eq{eq:tadpole0} leads to great simplifications for
numerical applications with more elaborate approximations.

Now, we proceed with the contributions from the three-point functions
in \eqref{eq:twopointflow_Gdot}. First, we use the symmetric occurrence
of the propagators in the loop. This allows us to pull out the
$\tau$-derivative in front of both propagators, schematically this
reads $G_\tau\partial_\tau G_\tau \simeq (1/2) \partial_\tau (G^2)$ in
the integral. Then we rewrite, similarly to the tadpole case, the
three-point function contribution as a total $\tau$\nobreakdash-derivative and
$\partial_\tau\Gamma_\tau^{(3)}$-terms,
\begin{align}\nonumber
  &\frac\imag2\int_{\contourC(z_1,\dots,z_4)} \Biggl( \partial_\tau 
    \left[
    \Gamma^{(3)}_{\tau, xz_1z_2} G_{\tau, z_2z_4}
    \Gamma^{(3)}_{\tau, z_3z_4y} G_{\tau, z_1z_3} \right]\\[1ex] \nonumber
  &\hspace{1cm} - 
    ( \partial_\tau\Gamma^{(3)}_{\tau, xz_1z_2})\, G_{\tau, z_2z_4}
    \Gamma^{(3)}_{\tau, z_3z_4y} G_{\tau, z_1z_3}\\[1ex]
  &\hspace{1cm} - 
    \Gamma^{(3)}_{\tau, xz_1z_2} G_{\tau, z_2z_4}
    (\partial_\tau\Gamma^{(3)}_{\tau, z_3z_4y} )\,G_{\tau, z_1z_3}\Biggr)\,.
	\label{eq:partintflow}
\end{align}
In $\partial_\tau\Gamma^{(3)}$, we only have to consider the parts
proportional to $\delta(\tau -x^0)$ and $\delta(\tau -y^0)$, since the
local terms are absent. Moreover, the term proportional to
$\delta(\tau -y^0)$ vanishes as the cutoff time in this term,
$\tau=y^0$, is smaller than the external time $x^0$. The first line is
a total $\tau$\nobreakdash-derivative and can be trivially integrated. In
summary, the $\tau$\nobreakdash-integration of \eq{eq:partintflow} leads us to
\begin{align}\nonumber 
  &\frac\imag2 \int_{\contourC(z_1,\dots,z_4)\leq x^0}
    \left[\Gamma^{(3)}_{xz_1z_2}  -
    \Delta\Gamma^{(3)}_{xz_1z_2} \right]\, G_{z_2z_4}
    \Gamma^{(3)}_{z_3z_4y} G_{z_1z_3} \\
  & =\frac\imag2 \int_{\contourC(z_1,\dots,z_4)\leq x^0}
    \Gamma^{(3)}_{t_0,xz_1z_2}   G_{z_2z_4}
    \Gamma^{(3)}_{z_3z_4y} G_{z_1z_3}\,.
    \label{eq:GenGap}
\end{align}
\Eq{eq:GenGap} is (the right hand side of) the gap equation with an
initial action $\Gamma_{t_0}$.  In the present work, we use the
classical action $S$ in \eqref{eq:action} as the initial action. Thus,
$\Gamma^{(3)}_{t_0} = S^{(3)}$ and $\Gamma^{(2)}_{t_0} = S^{(2)}$. 
Then, \eq{eq:GenGap} reduces to the
familiar gap equation
\begin{align}
	\Gamma^{(2)}_{xy} - S^{(2)}_{xy} = - \frac{\imag\lambda}{2} 
  \int\displaylimits_{\mathrlap{\contourC(z_3,z_4) \leq x^0}}
  \Gamma^{(3)}_{z_3z_4y} G_{xz_3} G_{xz_4}\,.
	\label{eq:gapintegratedflow}
\end{align}
For the truncation in the present work with $\Gamma^{(3)}=S^{(3)}$, we
directly obtain the integrated flow of our truncation
\eqref{eq:integrated_flow} from \eqref{eq:gapintegratedflow}.

\subsection{Gap equation in the
  \texorpdfstring{$\phi^4$}{phi\textasciicircum
    4}-theory}\label{eq:gap4}

The lack of local terms in the $n$-point functions for the
$\phi^3$-theory does not hold in general theories. While a full
analysis is beyond the scope of the present work, the generic
structure can be elucidated within the $\phi^4$-theory. From
\eq{eq:tadpole}, we get a generalised version of \eqref{eq:tadpole0}
that takes the local contributions into account
\begin{align}\nonumber 
  &	\frac12\int_{{\contourC(z_1,z_2)}\leq x^0} \Biggl\{
    \Gamma^{(4)}_{t_0,xyz_1z_2}  G_{z_1 z_2}  \\
  +&
     \,\Delta\Gamma^{(4)}_{\textrm{local}, xyz_1 z_2} G_{z_1 z_2}
     - \Delta\Gamma^{(4)}_{\textrm{local}, xyz_1 z_2} * G_{z_1 z_2}\Biggr\}\,.
\label{eq:tadpoleNot0}\end{align}
The second term stems from the total derivative. 
We also keep the restriction $x^0>y^0>t_0$ already used in the
previous \sec{sec:gap3}. Note also that the term at the initial cutoff
$\tau=t_0$ still vanishes trivially despite $\Gamma_{t_0}^{(4)}\neq 0$
as $G_{t_0}(x,y)=0$. We emphasise that $\Delta\Gamma^{(4)}* G$ is a
short hand notation for
$\int_{x^0-\epsilon}^{x^0+\epsilon} \partial_\tau \Delta\Gamma^{(4)} *
G$.

Before we proceed with the computation, we discuss the diagrammatic
relevance of the two terms in \eq{eq:tadpoleNot0}: the first term can
already be identified with a diagram in the gap equation, the tadpole
with an initial four-point function. Naturally, it is absent in the
$\phi^3$-theory. The second line is also non-vanishing, in
contradistinction to the $\phi^3$-theory, as the vertex correction
$\Delta\Gamma^{(4)}$ in the $\phi^4$\nobreakdash-theory contains local
parts. Indeed, one can show that all $n$-\nobreakdash point functions
contain local parts.  This general structure will be considered
elsewhere. Here, we elucidate this property with the simplest example
relevant for the present discussion, the fish diagrams with classical
vertices, which is the analogue of \eq{eq:triangle}. It is
self-consistently generated from the total derivative term in the
fourth derivative of \eq{eq:MasterFlowTotal} using the classical
vertices. Then, the vertex correction of the four-point vertex is
given by
\begin{align}\nonumber 
  \Delta\Gamma^{(4)}_{x y z_1 z_2} =
  \,
  \frac{\imag}{2} \lambda_4^2 \big[\big. & G^2(x,y) \deltac(x-z_1)
                                           \deltac(y-z_2) \\[1ex]\nonumber
  + \, & G^2(x,y) \deltac(x-z_2)\deltac(y-z_1) \\[1ex] \nonumber
  + \, & G^2(x,z_1) \deltac(x-y)\deltac(z_1-z_2) \big.\big]\\[1ex]
  + \, &\cdots \,, 
\label{eq:box} \end{align}
where $\lambda_4$ is the initial (classical) coupling of the
$\phi^4$\nobreakdash-theory. In \eq{eq:box}, the dots stand for other diagrams as
well as vertex corrections to the fish diagrams. The latter diagrams
are evidently local. Inserting \eq{eq:box} in the second line in
\eq{eq:tadpoleNot0}, we get flow contributions from the first two
explicit terms in \eq{eq:box}.

The term directly proportional to $\deltac(x^0-y^0)$ leads to
$2 \lim_{r\to 0}  \theta_r(\tau -x^0) \delta_r(\tau -x^0) = \delta(\tau -x^0)$. 
Hence, this contribution cancels in the second line of \eqref{eq:tadpoleNot0}.
In perturbation theory, this term potentially contributes to the 'double
tadpole', schematically given by
$ \deltac(x-y) G_{\textrm{cl},xz_1}^2 G_{\textrm{cl},z_1 z_1}$, where
an integral over $z_1$ is implied. Here, $G_\textrm{cl}$ is the
classical propagator. However, in our approach this term is contained completely in
the first term in \eq{eq:tadpoleNot0}. This is easily seen within a
perturbative expansion of the propagator. It also follows directly
from topological considerations of the diagrams in the gap equation: the
first term in \eqref{eq:tadpoleNot0} is a diagram in the gap equation
and the other diagrams do not generate the 'double tadpole' topology.

For the other terms, the $*$-product has to be evaluated. Thus, we
have to consider the dependence on $\tau$. Each of the propagators
$G(x,y)$ has to be multiplied with
${\theta_r(\tau - x^0) \theta_r(\tau - y^0)}$. We can ignore the
second $\theta$\nobreakdash-function as we only consider
$x^0>y^0$. This leads to
\begin{align}
  \partial_\tau \theta_r(\tau-x^0)^2 = 2
  \delta_r(\tau-x^0)\theta_r(\tau-x^0)\,.
\end{align}
The contraction of $G_{z_1 z_2}$ with the $\deltac(x^0_i-z^0_j)$ from
\eqref{eq:box} provides further $2\theta_r(\tau -x_0)$. Taking into account the 
factor $1/2$ from the vertex correction, cf. \eq{eq:box}, we arrive
at
\begin{align}\label{eq:Salm2}
  2 \lim_{r\to 0}  \theta_r(\tau -x^0)^2 \delta_r(\tau -x^0) =
  \frac23  \delta(\tau -x^0) \,.
\end{align}
The first term in the second line of \eqref{eq:tadpoleNot0}
contributes with a factor of one. The subtraction together with the
global factor $1/2$ gives a combinatorial factor $1/6$.
\begin{align}
  \frac{\imag}{2}\left[ \Delta\Gamma^{(4)}_{\textrm{local}} \cdot G
  - \Delta\Gamma^{(4)}_{\textrm{local}} * G\right]=
  \frac{\imag}{6}\lambda_4^2 G_{xy}^3 \,.
\label{eq:sunset}\end{align}
The right hand side of \eq{eq:sunset} is nothing but the sunset graph
with full propagators, and the combinatorial factor is the correct one
that can be deduced from perturbation theory. Notably, the vertex
correction $\Delta\Gamma^{(4)}$ does not give contributions to the
'double tadpole' diagram in perturbation theory. This is another
unique property that stems from the present local and causal
regulator.

In summary, we have seen how the local and causal structure can be
used for computing the corrections. We also remark that the different
powers of $\theta$\nobreakdash-functions can be included
systematically.  For example, one easily obtains
\begin{align}\label{eq:Salm}
  \lim_{r\to 0}f\!\left[\theta_r(\tau-x^0)\right ]
  \delta_r(\tau-x^0) = \delta(\tau-x^0)
  \int_0^1 dx\, f[x]  \,.
\end{align}
A detailed analysis is beyond the scope of the present work as no
local terms are present in the $\phi^3$-theory.  More details will be
presented elsewhere.

Now, we proceed with the contributions of the three-point functions in
\eqref{eq:twopointflow_Gdot}. For $\phi=0$, they are absent in the
$\phi^4$-theory but are present for $\phi\neq 0$ or in a theory with
additional microscopic $\phi^3$ vertices. The general expression is
obtained by simply adding the local terms to \eq{eq:partintflow} that
where absent in the $\phi^3$-theory. We arrive at
\begin{align}\nonumber 
  &\frac\imag2 \int_{\contourC(z_1,\dots,z_4)\leq x^0}\Biggl\{
    \Gamma^{(3)}_{t_0,xz_1z_2}   G_{z_2z_4}
    \Gamma^{(3)}_{z_3z_4y} G_{z_1z_3} \\[1ex]\nonumber 
  &\hspace{1.3cm}+
    \Delta\Gamma^{(3)}_{\mathrm{local},xz_1z_2} \left[ G_{z_2z_4} 
    \Gamma^{(3)}_{z_3z_4y} G_{z_1z_3} \right] \\[1ex]
  &\hspace{1.3cm}- \Delta\Gamma^{(3)}_{\mathrm{local},xz_1z_2}*
    \left[ G_{z_2z_4} 
    \Gamma^{(3)}_{z_3z_4y} G_{z_1z_3} \right]\Biggr\}\,.
    \label{eq:GenGap4}
\end{align}
The first line in \eq{eq:GenGap4} can already be identified with the
respective term in the gap equation.  Local terms for the correction
of the three-point function are only generated from the
$\phi^4$-interaction. We also remark that \eq{eq:GenGap4} 
contains part of the squint diagram in the gap equation of the mixed
theory with classical (initial) three- and four-point vertices
$S^{(3)}$ and $S^{(4)}$. The rest of the squint diagram is generated
by the respective vertex corrections to $\Delta\Gamma^{(4)}$ in the
tapole \eq{eq:tadpoleNot0}. The squint diagram is the last missing
diagram in the gap equation for the mixed theory.

Putting everything together, we are lead to a remarkable novel result:
a one-loop exact functional relation for the two-point function valid
for general theories. We emphasise that other functional relations for
correlation functions do not have a generic one loop form, for example
the Dyson-Schwinger equation for the propagator is only one-loop exact
in the $\phi^3$-theory, while it is two-loop exact in the
$\phi^4$-theory. The integrated flow of the two-point function for
$x^0\geq y^0$ is given by 
\begin{widetext}
\begin{align}\nonumber
 	\Gamma^{(2)}_{xy} - \Gamma^{(2)}_{t_0,xy} =&\,
    \frac12\int_{{\contourC(z_1,z_2)}\leq x^0}
    \Biggl\{
    \Gamma^{(4)}_{t_0,xyz_1z_2}  G_{z_1 z_2}  +
     \left( \Delta\Gamma^{(4)}_{\textrm{local}, xyz_1 z_2}
     - \Delta\Gamma^{(4)}_{\textrm{local}, xyz_1 z_2} *\right)  G_{z_1 z_2}
     \Biggr\}\\[1ex]
   & \hspace{-2cm}+\frac\imag2 \int_{\contourC(z_1,\dots,z_4)\leq x^0}\Biggl\{
    \Gamma^{(3)}_{t_0,xz_1z_2} \, G_{z_2z_4}
    \Gamma^{(3)}_{z_3z_4y} G_{z_1z_3} +
    \left( \Delta\Gamma^{(3)}_{\mathrm{local},xz_1z_2} -
    \Delta\Gamma^{(3)}_{\mathrm{local},xz_1z_2}*\right)
    \left[ G_{z_2z_4} 
    \Gamma^{(3)}_{z_3z_4y} G_{z_1z_3} \right] \Biggr\}\,.
     	\label{eq:fullintegratedflow4}
\end{align}
\end{widetext}
The first terms on the right hand side in the first and second line
are the one loop diagrams in the gap equation, the second terms
generate two-loop terms. As mentioned before, this one-loop exact
relation is the general result for generic theories independent of the
initial action. Using the same reasoning,
$\partial_\tau \Gamma^{(n)}_\tau$ can be integrated analytically for
any $n$. We close with the remark that this is only seemingly in
contradiction with the proof that such one-loop exact functional
relations do not exist in \cite{Litim:2002xm}. The present approach
implicitly escapes one of the presuppositions there (no integral over
parameters such as the cutoff time) via \eq{eq:Salm}.

\section{Initial conditions}
\label{app:init}

We employ a mixed position and momentum space, where only the spatial
coordinates are replaced by momenta.

In case of using the integral equation \eqref{eq:dysoneq}, the initial
conditions are encoded in the solution of the free equation of
motion $\bar{G}$, or $\bar{F}$ and $\bar{\rho}$
\begin{align}\nonumber
  \bar{F}(t,t';\vec{p}) &= \frac{1}{\omega_{\vec{p}}}
                          \left(\frac{1}{2}+f_{0}(\vec{p})\right)\cos\big[
                          \omega_{\vec{p}}(t-t')\big]\,, \\[1ex]
  \bar{\rho}(t,t';\vec{p}) &= \frac{1}{\omega_{\vec{p}}}\sin\big[
                             \omega_{\vec{p}}(t-t')\big]\,.
\end{align}
Here, $f_{0}(\vec{p})$ is the occupation number and
$\omega_{\vec{p}} = \sqrt{\vec{p}^2+m_0^2}$ the dispersion relation at
initial time.

For the differential equation approach \eqref{eq:diffeq}, we have to
provide values for $F$ and $\rho$ and their first derivatives at the
initial time. In analogy to the previous case, we use for the
statistical two-point function
\begin{align}\nonumber 
  F(t,t';\vec{p})\vert_{t=t'=t_0}
  &= \frac{1}{\omega_{\vec{p}}}
    \Big(\frac{1}{2}+f_{0}(\vec{p})\Big)\,,
  \\[1ex]\nonumber
  \partial_{t}F(t,t';\vec{p})\big\vert_{t=t'=t_0}
  &= 0\,,
  \\[1ex]
  \partial_{t}\partial_{t'}F(t,t';\vec{p})\vert_{t=t'=t_0}
  &=
    \omega_{\vec{p}} \Big(\frac{1}{2}+f_{0}(\vec{p})\Big)\,,
	\label{eq:Fpp_initial}
\end{align}
and the spectral function
\begin{align}\nonumber
  \rho(t,t';\vec{p})\vert_{t=t'=t_0}
  &= 0\,, \\[1ex]\nonumber
  \partial_{t}\rho(t,t';\vec{p})\big\vert_{t=t'=t_0} &= 1\,,\\[1ex]
  \partial_{t}\partial_{t'}\rho(t,t';\vec{p})\vert_{t=t'=t_0} &= 0\,.
\end{align}
For the results shown in \ref{sec:results}, we used 
\begin{align}
  f_{0}(\vec{p}) =
  \frac{\widetilde{N}}{\widetilde{\lambda}}
  \theta(m_{0}-\vert\vec{p}\vert)\,,
	\label{eq:appinit}
\end{align}
with $\widetilde{N}=100$. The dimensionless coupling of the the
three-point function is given by
$\widetilde{\lambda}=\lambda/m_0^2=0.01$.

\section{Total energy}
\label{app:T00}

The total energy is given by $\langle T_{00}\rangle$ and the
expectation value of the energy-momentum tensor is obtained from the
effective action as
\begin{align}\label{eq:Tmunu}
  \langle T_{\mu\nu}(x)\rangle  =  \left. \frac{2}{\sqrt{-g(x)}}
  \frac{\delta \Gamma[\phi,g]}{\delta g^{\mu\nu}(x)}
  \right|_{g^{\mu\nu}=\eta^{\mu\nu}}
  \,.
\end{align}
In \eq{eq:Tmunu}, the metric $g^{\mu\nu}$ is identified with the
Minkowski metric $\eta^{\mu\nu}$. The flow of
$\langle T_{\mu\nu}\rangle$ can be derived from the metric variation
of $\partial_\tau\Gamma_\tau$ and will be discussed elsewhere.  Here,
we follow closely derivations also found in the 2PI framework,
e.g.~\cite{Berges2017}, which allows us to discuss the diagrammatical
consistency of our truncation.

Concentrating on $\langle T_{00}\rangle $, we use that $T_{00}$ is
derived from \eq{eq:Tmunu} by substituting $\Gamma$ by the classical
action \eq{eq:action}. This leads us to
\begin{align}
  T_{00}(x) =  \partial_0 \varphi \partial_0\varphi - g_{00}
  \left( \frac12 \partial_\mu
  \varphi \partial^\mu\varphi -\frac12 m^2_0\varphi^2 -
  \frac{\lambda}{3!}\varphi^3\right) \,.
\end{align}
For its expectation value, we use that (see e.g.\cite{Pawlowski:2005xe})
\begin{align}\label{eq:funRepExp}
  \left\langle \prod_{i=1}^n \varphi(x_i)\right\rangle =  \prod_{i=1}^n\left[
  \int\displaylimits_{\mathrlap{\contourC(z_i)}}  G(x_i , z_i)\frac{\delta }{\delta \phi(z_i) }+
  \phi(x_i)\right]_{\phi_\textrm{EoM}} \,,
\end{align}
with the full mean field-dependent propagators $G[\phi](x,y)$ and the
mean field $\phi=\langle \varphi\rangle $
being evaluated on the equations of motion (EoM). In the present case,
we have $\phi_\textrm{EoM}=0$. Thus, $\langle T_{00}\rangle$
reduces to
\begin{widetext}
\begin{align}
  \langle T_{00}(x)\rangle =
  \frac12 \lim_{y \to x} \Bigl[  \partial_{x^0} \partial_{y^0} G(x,y)  
  + \left( -\partial_{\vec{x}}^2 +m_0^2 \right) G(x,y)\Bigr]
  +  \frac{\imag\lambda}{3!} \int
  \displaylimits_{\mathrlap{\contourC(z_1,z_2,z_3)}}
  \Gamma^{(3)}(z_1,z_2,z_3) \prod_{i=1}^3 G(x,z_i) \,,   
\label{eq:EMT-Exp}\end{align}
where we have used
\begin{align}
  \frac{\delta}{\delta\phi(z_1)} G(x,y) =\imag
  \int\displaylimits_{\mathrlap{\contourC(z_2,z_3)}} 
  G(x,z_2) \Gamma^{(3)}(z_1,z_2, z_3) G(z_3,y)\,.
\end{align}
The last term in \eq{eq:EMT-Exp} is the vacuum sunset term. Its one
loop subgraph is related to the gap equation (Dyson-Schwinger
equation)
\begin{align}
  \Gamma^{(2)}(x,y) =
  S^{(2)}(x,y)   - \frac{\imag\lambda}{2} \int
  \displaylimits_{\mathrlap{\contourC(z_1,z_2)}}
  \Gamma^{(3)}(y,z_1,z_2)
  G(z_1,x) G(z_2,x)\,, 
\label{eq:gap}\end{align}
which leads us to
\begin{align} \frac{\imag\lambda}{3!}
    \int\displaylimits_{\mathrlap{\contourC(z_1,z_2,z_3)}} 
    \Gamma^{(3)}(z_1,z_2,z_3) \prod_{i=1}^3 G(x,z_i) = \frac13 \int\displaylimits_{\mathrlap{\contourC(z)}}  \left[  
      S^{(2)}(x,z) -\Gamma^{(2)}(x,z) \right]  G(z,x)\,.
      \label{eq:DSE}
\end{align}
\end{widetext}
The second term is proportional to an irrelevant constant,
$(\Gamma^{(2)}\cdot G)(x,y)=\imag\,\delta_{\cal C}(x-y)$, while the
first one simply changes the prefactors of the first and second term
on the right hand side of \eq{eq:EMT-Exp}. Fourier transforming the
spatial coordinates, we arrive at
\begin{align}\nonumber
  \langle T_{00}(t)\rangle =
  &\, 
   \frac56 \lim_{t\to t'}    \partial_{t} \partial_{t'}  \int_{\vec p}
    G(t,t';{\vec p}) \\[1ex]
  &+ \frac16   \int_{\vec p}\left( {\vec p}^2 +m_0^2\right) 
    G(t,t;{\vec p})\,.
   \label{eq:EMT-Exp-Reduced}
\end{align}
Note that the explicit occurrence of $\Gamma^{(3)}$ has dropped out in
\eq{eq:EMT-Exp-Reduced}. This trivially ensures the self-consistency
of \eq{eq:EMT-Exp-Reduced} with the approximation of $\Gamma^{(3)}$
used in the computation of the propagator. We also remark that
\eq{eq:EMT-Exp-Reduced} readily extends to all components of the
expectation value of the energy momentum tensor
$\langle T_{\mu\nu}\rangle$.

\section{Numerical details}
\label{app:numerics}

The initial conditions for the results shown in this section can be
found in appendix \ref{app:init}. The parameters used are the same as
in the rest of the paper (cf. \eqref{eq:appinit} and below).

To solve the integro-differential equation \eqref{eq:diffeq}, we use a
symmetric discretisation of the second-order time-derivative. This
allows us to use the simple explicit Euler method to compute the
solution. More details can be found e.g. in \cite{Berges2005}. All
integrals are computed using a trapezoidal rule.

\begin{figure}[t]
	\centering
	\includegraphics{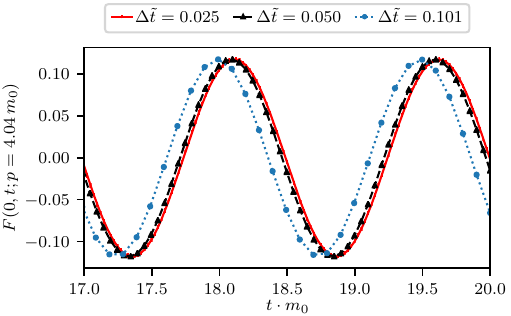}
	\caption{Time evolution of the statistical propagator obtained
          using the differential solver for various time resolutions
          ${\Delta \tilde{t} = \Delta t\cdot m_0}$. The curves get
          closer to each other for decreasing time-step size.}
	\label{fig:late_time_F_diff_res}
\end{figure}

In \fig{fig:late_time_F_diff_res}, we show the solution for
statistical propagator $F(0,t,p=4.04m_0)$ for three different
time-step sizes $\Delta \tilde{t} = \Delta t \cdot m_0$. For a decreasing
step size, the curves get closer to each other. We do not observe an
instability for the step sizes used.

\begin{figure}[t]
	\centering
	\includegraphics{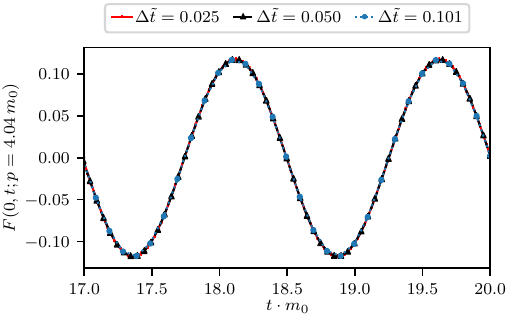}
	\caption{Time evolution of the statistical propagator obtained
          using the explicit solver for various time resolutions
          ${\Delta \tilde{t} = \Delta t \cdot m_0}$. The curves are
          perfectly on top of each other. This demonstrates the faster
          convergence of the explicit solver as compared to the
          differential one.}
	\label{fig:late_time_F_exp_res}
\end{figure}

In \fig{fig:late_time_F_exp_res}, we show $F(0,t,p=4.04m_0)$ as
obtained from the integral equation \eqref{eq:dysoneq} for the same
times as in \fig{fig:late_time_F_diff_res} and the same step
sizes. We observe that the curves are perfectly on top of each other
for all shown step sizes. Thus, the explicit solver obtained from the
discretisation of the integral equation \eqref{eq:dysoneq} converges
faster than the differential one. Details on how to solve
\eqref{eq:dysoneq} as an explicit equation will be given elsewhere.

\begin{figure}[b]
	\centering
	\includegraphics{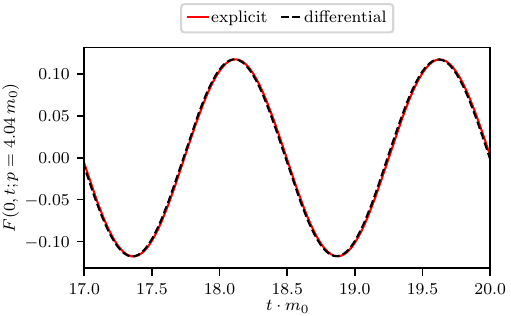}
	\caption{Time evolution of the statistical propagator obtained
          using the differential and explicit solver for
          ${\Delta \tilde{t} = \Delta t \cdot m_0 = 0.025}$. Both
          solvers agree for small enough time-step size.}
	\label{fig:late_time_F_diff_vs_exp}
\end{figure}

What remains to be shown is that both methods agree for sufficiently
small step sizes. This can be seen from
\fig{fig:late_time_F_diff_vs_exp} where we compare both solvers for
the smallest available step size. The same agreement is obtained
comparing the largest one for the explicit solver with the smallest
one of the differential solver.

Taking a closer look at \fig{fig:late_time_F_diff_vs_exp}, we see
that the black dashed line is still slightly shifted compared to the
red one. This again demonstrates the faster convergence of the
explicit solver.

The faster convergence comes at the price of an additional time
integral that has to be computed. However, the differential solver
requires a smaller time-step size to produce results of the same
accuracy. For the results shown in this section for
$\Delta \tilde t=0.025$ for the differential solver and
$\Delta\tilde t=0.101$ for the explicit solver, the runtime of both
solvers is comparable (same order of magnitude).

\bibliography{../bib_master}

\end{document}